\documentclass[ aps, pra,
 amsmath,amssymb,
 citesuperscript,
 superscriptaddress,
 reprint,%
 showkeys,
 twocolumn,
]{revtex4-1}

\usepackage{amsmath,amssymb,amsfonts}
\usepackage{algorithmic}
\usepackage{graphicx}
\usepackage{subfigure}
\usepackage[sort&compress]{natbib}
\usepackage{booktabs}
\usepackage{tabularx}
\usepackage{longtable}
\usepackage{multirow}
\usepackage{float}
\usepackage{textcomp}
\usepackage{xcolor}
\usepackage[version=4]{mhchem}
\usepackage{enumerate}
\usepackage{enumitem}
\usepackage{siunitx}
\setcitestyle{super,bracket}
\usepackage{xr}
\begin{document}

%\preprint{}

\title{Solid-Electrolyte Interphase During Battery Cycling: Theory of Growth Regimes} %Title of paper
\author{Lars von Kolzenberg}
 \affiliation{Institute of Engineering Thermodynamics, German Aerospace Center (DLR) Pfaffenwaldring 38-40, 70569 Stuttgart (Germany)}
\affiliation{Helmholtz Institute Ulm (HIU), Helmholtzstraße 11, 89081 Ulm (Germany)}
\author{Arnulf Latz}
\affiliation{Institute of Engineering Thermodynamics, German Aerospace Center (DLR) Pfaffenwaldring 38-40, 70569 Stuttgart (Germany)}
\affiliation{Helmholtz Institute Ulm (HIU), Helmholtzstraße 11, 89081 Ulm (Germany)}
\affiliation{Ulm University (UUlm), Albert-Einstein-Allee 47, 89081 Ulm (Germany)}
\author{Birger Horstmann}
 \email{birger.horstmann@dlr.de}
\affiliation{Institute of Engineering Thermodynamics, German Aerospace Center (DLR) Pfaffenwaldring 38-40, 70569 Stuttgart (Germany)}
\affiliation{Helmholtz Institute Ulm (HIU), Helmholtzstraße 11, 89081 Ulm (Germany)}
\affiliation{Ulm University (UUlm), Albert-Einstein-Allee 47, 89081 Ulm (Germany)}

\date{\today}

\begin{abstract}
The capacity fade of modern lithium ion batteries is mainly caused by the formation and growth of the solid-electrolyte interphase (SEI). 
Numerous continuum models support its understanding and mitigation by studying SEI growth during battery storage.
However, only a few electrochemical models discuss SEI growth during battery operation. 
In this article, we develop a continuum model, which consistently captures the influence of open circuit potential, current direction, current magnitude, and cycle number on the growth of the SEI. 
Our model is based on the formation and diffusion of neutral lithium atoms, which carry electrons through the SEI. Recent short- and long-term experiments provide validation for our model. We show that SEI growth is either reaction, diffusion, or migration limited. For the first time, we model the transition between these mechanisms and explain empirically derived capacity fade models of the form $\Delta Q\propto t^\beta$ with $0 \leq \beta \leq 1$. Based on our model, we identify critical operation conditions accelerating SEI growth.
\end{abstract}


\maketitle %\maketitle must follow title, authors, abstract
% The preceding line is only needed to identify funding in the first footnote. If that is unneeded, please comment it out.

\section{Introduction}

Lithium-ion batteries constitute the state of the art portable energy storage device as they provide high energy densities and long cycle lives.
Increased battery lifetime and safety would promote the emergence of electromobility. However, continued capacity fade of lithium-ion batteries remains as important challenge. The main cause of this capacity fade is the formation and growth of a solid-electrolyte interphase (SEI) on the graphitic anode \cite{Peled1979,Ploehn2004,Vetter2005,Staniewicz2005}. Understanding the structure, composition, and continued growth of the SEI is thus key to extend battery life, improve battery safety, and develop new high-energy electrodes.

The SEI is a thin layer, which forms during the first charging cycle, when the anode potential falls below the electrolyte reduction potential \cite{Goodenough2010,Horstmann2019,Wang2018}. Electrolyte molecules react with electrons and lithium ions forming a nanometer thick layer of solids on the anode surface \cite{Peled2017,Aurbach2000}. Although this layer protects the electrolyte from low anode potentials in subsequent cycles, the SEI continues to grow and consumes lithium ions in the process. 

Different experiments have revealed that the SEI exhibits a dual-layer structure with a dense inner layer and a porous outer layer. Anorganic compounds like $\ce{LiF},\ce{Li_2CO_3}$ and $\ce{Li_2O}$ build up the inner layer and organic compounds like $\ce{Li_2EDC}$ build up the outer layer \cite{lu2014chemistry,Wang2018,Peled1997,Aurbach1999,Winter2009,Xu2004,Agubra2014,An2016,Lu2011,Shi2012,Xu2014}.
Recent cryogenic electron microscopy measurements \cite{Huang2019,Boniface2016a} give evidence that the different layers grow next to each other on the particle surface. Some graphite particles are covered in a slowly growing dense SEI, while others are surrounded by a fast growing porous SEI. The experimental characterization of the underlying transport and reaction mechanisms is impeded by small length scales, air sensitivity, and the chemical variety of the SEI. 

Electrochemical models give valuable complementary insights to reveal the transport and reaction processes within the SEI. It is well-established that transport processes limit SEI growth during long-term battery storage. Transport limitations lead to a capacity fade proportional to the square root of elapsed time, i.e. $\sqrt{t}$. Different mechanisms are proposed to explain this behavior \cite{Horstmann2019,Reniers_2019}, including solvent diffusion \cite{Tang2012a,Ploehn2004,Pinson2012,Tang2011,Tang2012b,Single2016,Single2017,Tahmasbi2017,Hao2017,Roder2016,Single2018}, electron conduction \cite{Staniewicz2005,Christensen2004,Colclasure2011,Roder2017,Das2019,Pinson2012, Single2016,Single2017}, electron tunneling \cite{Tang2012b,Li2015,Single2018}, and the diffusion of neutral lithium interstitial atoms \cite{Single2018, Soto2015,Shi2012}.
In a comparative study of these mechanisms, Single et al. \cite{Single2018} identify neutral lithium diffusion as likely transport mechanism, because it explains the state of charge dependence of the extensive storage experiments of Keil et al. \cite{Keil2016a,Keil2016b}. 

During battery operation, however, the external conditions, e.g., charging rate and depth of discharge, strongly influence the SEI growth rate. Several papers analyze the resulting capacity fade with empirical formulas \cite{Kabitz2013,Schmalstieg2014,Groot2015,Hahn2018,Severson2019,Attia2020}. These approaches nicely agree with experimental measurements, but do not give further insights into underlying growth mechanisms. Physics-based models for SEI growth during battery operation remain scarce and rely on solvent diffusion \cite{Ekstrom2015}, electron conduction \cite{Das2019}, or electron tunneling \cite{Li2015} as charge transport mechanism.

In a recent joint experimental and theoretical work, Attia and Das et al. \cite{Attia2019,Das2019} investigate the influence of current, voltage and cycle number on SEI growth. Attia et al. \cite{Attia2019} measure the differential capacity $\text{d}Q/\text{d}V$ during intercalation and deintercalation of carbon black. They isolate the SEI contribution by comparing the second cycle with a high SEI contribution to a later baseline cycle with hardly any SEI contribution. Thereby, they show an asymmetry in SEI growth: During charging the SEI grows faster than during discharging. Das et al. \cite{Das2019} model this asymmetry by assuming that the SEI is a mixed ionic electronic conductor. In this model, the SEI conductivity depends on the concentration of lithium ions inside the SEI. The lithium ion concentration inside the SEI and thereby the SEI formation current is high during charging and low during discharging. However, there are some inconsistencies in the modeling approach. \textit{First}, recent models show that the SEI is a single-ion solid electrolyte \cite{Single2019}. Therefore, the lithium ion concentration inside the SEI should remain constant due to charge conservation. \textit{Second}, the modeled conduction of electron and lithium leads to counterpropagating fluxes. Thus, SEI formation should be fully suppressed during deintercalation. \textit{Third}, the proposed mechanism of electron conduction disagrees with the electrode potential dependence of SEI growth observed in long-term storage experiments \cite{Keil2016a,Keil2016b}. Instead, the diffusion of radicals can explain these observations \cite{Single2018}.

In this paper, we discuss a consistent understanding of transport through the SEI and the dependence of SEI growth on operating conditions. The model consistently links the short-term behaviour measured in the experiments of Attia et al. \cite{Attia2019} with the long-term storage behavior measured by Keil et al. \cite{Keil2016a,Keil2016b}. For the first time, our approach shows the transition between different growth regimes, achieved by the coupling of the formation reaction and diffusion process of neutral lithium interstitial atoms in the SEI.

We present our model development in section \ref{s:Theory} and explain our implementation in section \ref{s:Methods}. In section \ref{s:Results}, we validate the simulation with short- and long-term experiments of Attia et al. \cite{Attia2019} and show results for very long times. We make use of our model in section \ref{s:Discussion} to analyze the influence of operating conditions on SEI growth with a focus on time dependence. Finally, section \ref{s:Conclusion} summarizes the key findings of this work.

\section{Theory}
\label{s:Theory}

\begin{figure}[t] 
 \centering
 \includegraphics[width=8.4 cm]{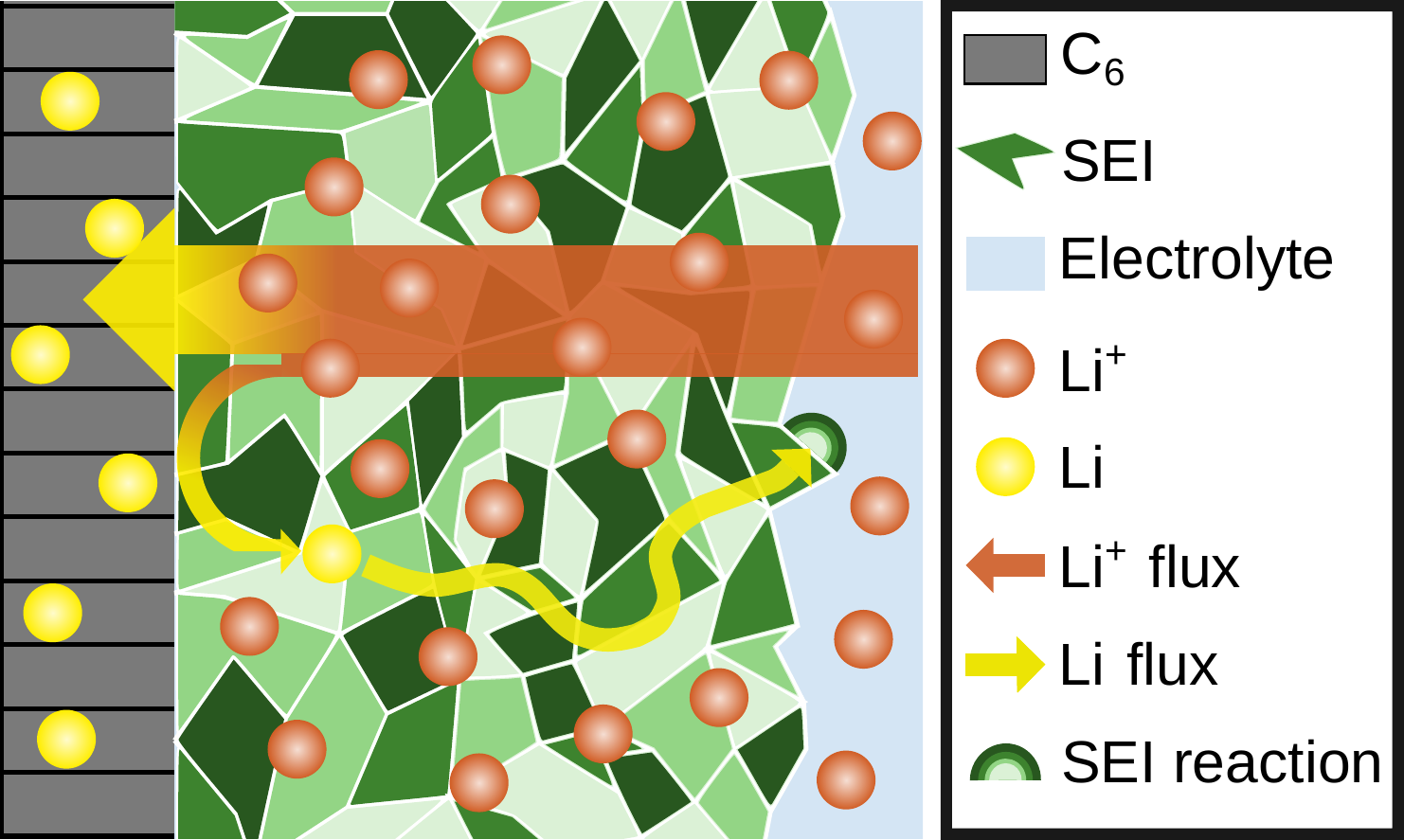}
 \caption{Schematic of the transport and reaction mechanisms in the SEI during battery charging. Neutral Lithium atoms form at the electrode and move to the SEI-electrolyte interface by interstitial diffusion and electron hopping. Then they react with electrolyte and form fresh SEI. Lithium ions migrate through the SEI.}
 \label{fig:SEI_growth}
\end{figure}

In this section, we present our theory for SEI growth based on the concept depicted in figure \ref{fig:SEI_growth}. 
At the electrode-SEI interface, lithium ions $\ce{Li}^+_\text{SEI}$ from the SEI react with electrons $\ce{e}^-$ from the electrode. The resulting lithium atoms either intercalate into the electrode in the form of $\ce{Li_xC_6}$ (see equation \ref{eq:SEI_intercalation_reaction}) or remain as neutral lithium interstitial atoms $\ce{Li}^0$ inside the SEI (see equation \ref{eq:SEI_interstitial_reaction}), 
\begin{align}
\label{eq:SEI_intercalation_reaction}
x\ce{Li}^+_\text{SEI} +x\ce{e}^- + \ce{C}_6& \rightleftharpoons \ce{Li_xC_6}, \\
\label{eq:SEI_interstitial_reaction}
\ce{Li}^+_\text{SEI} +\ce{e}^- & \rightleftharpoons \ce{Li}^0 .
\end{align} 
The lithium interstitial atoms $\ce{Li}^0$ subsequently move through the SEI to the SEI-electrolyte interface, where they immediately react and form new SEI. According to reaction equations \ref{eq:SEI_intercalation_reaction} and \ref{eq:SEI_interstitial_reaction}, the overall measured current of electrons $j$ consists of the intercalation $j_\text{int}$ and the SEI formation current $j_\text{SEI}$,
\begin{equation}
\label{eq:j_ges}
j=j_\text{int}+j_\text{SEI} .
\end{equation}
In section \ref{ss:Diffusion}, we discuss the equations for transport of neutral lithium atoms $\ce{Li^0}$. Afterwards, in sections \ref{ss:Intercalation} and \ref{ss:Formation}, we derive an expression for the kinetics of lithium intercalation and neutral lithium atom formation. Finally, we combine the formation and transport currents of lithium atoms to obtain an expression for the SEI growth rate $j_\text{SEI}$ and the resulting SEI thickness $L_\text{SEI}(t)$ in section \ref{ss:Expression}.

\subsection{Transport of neutral lithium atoms}
\label{ss:Diffusion}
We divide the electron transport from the electrode-SEI to the SEI-electrolyte interface into two contributions. First, the electrons tunnel a distance $L_\text{tun}$ into the SEI and react to $\ce{Li^0}$, according to equation \ref{eq:SEI_interstitial_reaction}. Second, the electrons move as neutral lithium interstitial atoms $\ce{Li^0}$ to the SEI-electrolyte interface. 
We account for the tunneling process by introducing an apparent SEI thickness 
\begin{equation}
L_\text{app}=L_\text{SEI}-L_\text{tun}.
\end{equation} 
Electrons could either move together with a neutral lithium atom or hop between lithium ions. For both cases, we use dilute solution theory \cite{newman2012electrochemical} to model the transport current $j_\text{SEI}$,
\begin{equation}
\label{eq:Dilute_Solution_1}
j_\text{SEI}=\underbrace{\vphantom{\frac{1}{2}}-z_{\text{e}^-}FD\nabla c_\text{Li}}_\text{diffusion}-\underbrace{\frac{z_{\text{e}^-}^2D F^2}{RT}c_\text{Li} \nabla \phi}_\text{electromigration}
\end{equation}
with the diffusion coefficient $D$ and the concentration $c_\text{Li}$ of neutral lithium atoms inside the SEI.
Here, $F$ is Faraday's constant, $R$ the universal gas constant, and $T$ the temperature. The electromigrative part of the flux describes electron transport due to an external electric field and depends on the valency of an electron $z_{\text{e}^-}=-1$ and the electrical potential $\phi$ in the SEI. 

We linearly approximate the gradients along the diffusion-migration path $L_\text{tun} \leq L \leq L_\text{SEI}$. We assume that electrons reaching the electrolyte are directly consumed to form new SEI, so that $c_\text{Li}(x=L_\text{SEI})=0$ \cite{Single2018}. Accordingly, the average concentration of lithium atoms inside the SEI is $\bar{c}_\text{Li} = c_\text{Li}(x=L_\text{tun})/2$. Using these assumptions and simplifications, we express the SEI current with equation \ref{eq:Dilute_Solution_2},
\begin{equation}
\label{eq:Dilute_Solution_2}
j_\text{SEI}=-DF\frac{c_\text{Li}(L_\text{tun})}{L_\text{app}}\left(1+\frac{F}{2RT} (\phi(L_\text{SEI})-\phi(L_\text{tun})) \right).
\end{equation}
Ohm's law gives an expression for the potential difference in equation \ref{eq:Dilute_Solution_2},
\begin{equation}
\label{eq:Ohms_Law}
\phi(L_\text{SEI})-\phi(L_\text{tun})=-\frac{L_\text{app}}{\kappa_{\text{Li}^+,\text{SEI}}}j_\text{int},
\end{equation}
with the lithium ion conductivity of the SEI $\kappa_{\text{Li}^+,\text{SEI}}$. Inserting equation \ref{eq:Ohms_Law} into \ref{eq:Dilute_Solution_2}, we obtain our final description of the diffusive-migrative electron current through the SEI,
\begin{align}
\label{eq:Dilute_Solution_3}
\begin{split}
j_\text{SEI}&=-\frac{c_{\text{Li}}(L_\text{tun})DF}{L_\text{app}}\left(1-\frac{F}{2RT}\frac{L_\text{app}}{\kappa_{\text{Li}^+,\text{SEI}}}j_\text{int} \right).
\end{split}
\end{align}

\subsection{Intercalation}
\label{ss:Intercalation}
We describe the intercalation current $j_\text{int}$ resulting from reaction \ref{eq:SEI_intercalation_reaction}, with a standard Butler-Volmer approach \cite{newman2012electrochemical,Latz2013,Bazant2013}
\begin{equation}
\label{eq:Butler-Volmer}
j_\text{int}=2j_0\sinh\left(\frac{F}{2RT}\eta_\text{int}\right).
\end{equation}
The consistent overpotential $\eta_\text{int}$ for reaction \ref{eq:SEI_intercalation_reaction} is defined by equation \ref{eq:Overpotential},
\begin{equation}
\label{eq:Overpotential}
\eta_\text{int}=\phi_\text{S}-U_0-\mu_{\text{Li}^+,\text{SEI}},
\end{equation}
with the electrode potential $\phi_\text{S}$, the open circuit voltage (OCV) $U_0$, and the electrochemical potential of lithium ions at the electrode-SEI interface $\mu_{\text{Li}^+,\text{SEI}}$.
Accordingly, intercalation overpotential $\eta_\text{int}$ and current $j_\text{int}$ are negative for intercalation and positive for deintercalation.
The consistent exchange current density $j_0$ defined by equation \ref{eq:Butler-Volmer_prefactor}, 
\begin{equation}
\label{eq:Butler-Volmer_prefactor}
j_0=j_{0,0}\sqrt{\frac{c_\text{s}}{c_\text{s,max}}},
\end{equation}
depends only on the lithium concentration inside the electrode $c_\text{s}$ relative to the maximum concentration $c_\text{s,max}$. We assume that the lithium ion concentration inside the SEI $c_{\text{Li}^+,\text{SEI}}$ is constant, because the SEI is a single-ion solid electrolyte with a fixed amount of charge carriers due to charge neutrality \cite{Single2018}. Thus, the exchange current density $j_\text{0,0}$ does not depend on $c_{\text{Li}^+,\text{SEI}}$. The concentration in the carbon black electrode $c_\text{s}$ changes over time according to equation \ref{eq:DAE1},
\begin{equation}
\label{eq:DAE1}
\frac{\text{d} c_\text{s}}{\text{d}{t}}=-\frac{A_\text{cb}}{F}j_\text{int},
\end{equation}
where $A_\text{cb}$ is the volume specific surface area of carbon black.

\subsection{Formation reaction of neutral lithium interstitials}
\label{ss:Formation}
SEI growth could be limited by two reactions, either neutral lithium interstitial formation at the electrode-SEI interface or electrolyte reduction at the SEI-electrolyte interface. Here, we present a simplistic model to enlighten the basic principles. Thus, we take into account only the kinetics of neutral lithium interstitial formation (see equation \ref{eq:SEI_interstitial_reaction}). We describe these reaction kinetics with an asymmetric Butler-Volmer approach \cite{newman2012electrochemical,Latz2013,Bazant2013},
\begin{equation}
\label{eq:Butler_Volmer_SEI}
j_\text{SEI}=j_\text{SEI,0} \cdot \left(e^{(1-\alpha_\text{SEI})\frac{F\eta_\text{SEI}}{RT}}-e^{-\alpha_\text{SEI}\frac{F\eta_\text{SEI}}{RT}} \right).
\end{equation}
We choose as asymmetry factor $\alpha_\text{SEI}=0.22$ in line with the density functional theory results of Li and Qi \cite{Li2019} and the microfluidic test cell measurements of Crowther and West \cite{Crowther2008}. The $\ce{Li^0}$ formation overpotential $\eta_\text{SEI}$ in equation \ref{eq:Butler_Volmer_SEI} follows from the reaction equation \ref{eq:SEI_interstitial_reaction} as
\begin{equation}
\label{eq:Overpotential_SEI}
\eta_\text{SEI}=\phi_\text{S}-\mu_{\text{Li}^+,\text{SEI}}+\mu_\text{Li}/F.
\end{equation}
We determine the chemical potential $\mu$ of the interstitial atoms with a dilute solution approach \cite{newman2012electrochemical},
\begin{equation}
\label{eq:chemical_potential}
\mu_\text{Li}=\mu_\text{Li,0}+RT\ln\left(\frac{c_{\text{Li}}}{c_{{\text{Li}},0}}\right).
\end{equation}
The chemical potential assumes its standard value $\mu_\text{Li,0}$ relative to lithium metal if the lithium atom concentration at the electrode-SEI interface $c_{\text{Li}}$ equals the reference concentration of $c_{{\text{Li}},0}=1\si{\mol\per\liter}$. 
The exchange current density $j_\text{SEI,0}$, 
\begin{equation}
\label{eq:Butler-Volmer_SEI_prefactor}
j_\text{SEI,0}=j_{\text{SEI},0,0}\left(\frac{c_{\text{Li}}}{c_{{\text{Li}},0}}\right)^{\alpha_\text{SEI}},
\end{equation}
depends on the interstitial concentration at the electrode $c_{\text{Li}}$, as we assume a constant lithium ion concentration inside the SEI.

We couple battery operation to $\ce{Li^0}$ formation by rephrasing equation \ref{eq:Butler_Volmer_SEI}. Combining equations \ref{eq:Butler_Volmer_SEI}-\ref{eq:Butler-Volmer_SEI_prefactor}, we obtain the following expression for the $\ce{Li}^0$ formation kinetics,
\begin{align}
\label{eq:Butler_Volmer_SEI_2}
j_\text{SEI}=j_\text{SEI,0,0} \cdot \left(\frac{c_{\text{Li}}(L_\text{tun})}{c_{{\text{Li}},0}}e^{(1-\alpha_\text{SEI})\tilde{\eta}_\text{SEI}}-e^{-\alpha_\text{SEI}\tilde{\eta}_\text{SEI}} \right).
\end{align}
The dimensionless potential jump for lithium atom formation, $\tilde{\eta}_\text{SEI}$, follows from combining equations \ref{eq:Overpotential_SEI}, \ref{eq:chemical_potential}, and \ref{eq:Overpotential}. This yields 
\begin{equation}
\label{eq:Overpotential_SEI_2}
\tilde{\eta}_\text{SEI}=\frac{F}{RT}(\eta_\text{int}+U_0+\mu_\text{Li,0}/F),
\end{equation}
as a function of the OCV $U_0$ and the intercalation overpotential $\eta_\text{int}$, which depends on intercalation current $j_\text{int}$ according to equation \ref{eq:Butler-Volmer}.

\subsection{SEI growth rates}
\label{ss:Expression}
So far, we derived expressions for the diffusive-migrative current through the SEI (equation \ref{eq:Dilute_Solution_3}) and the SEI growth based on the formation reaction of lithium atoms (equation \ref{eq:Butler_Volmer_SEI_2}). 
However, we do not know the current and voltage dependent concentration of lithium atoms $c_\text{Li}(L_\text{tun})$ inside the SEI. The two unknowns $j_\text{SEI}$ and $c_\text{Li}(L_\text{tun})$ are determined by the two equations \ref{eq:Butler_Volmer_SEI_2} and \ref{eq:Dilute_Solution_3}. This results in equation \ref{eq:SEI_current_final} for SEI growth ("+" for intercalation, "-" for deintercalation),
\begin{equation}
\label{eq:SEI_current_final}
j_{\text{SEI}}=-j_{\text{SEI,0,0}}e^{-\alpha_\text{SEI} \tilde{\eta}_\text{SEI}}\frac{1 \pm \frac{L_\text{app}}{L_\text{mig}}}{1\pm \frac{L_\text{app}}{L_\text{mig} }+\frac{L_\text{app}}{L_\text{diff}}}.
\end{equation}
Note that this is an implicit equation for $j_{\text{SEI}}$ as $\tilde{\eta}_\text{SEI}$ depends on $j_{\text{SEI}}$ through $\eta_\text{int}$ (see equation \ref{eq:Overpotential_SEI_2}).
In equation \ref{eq:SEI_current_final}, $L_\text{diff}$ and $L_\text{mig}$ are the critical thicknesses for diffusion and migration, respectively. They are defined by
\begin{align}
\label{eq:L_diff}
L_\text{diff}&=\frac{c_\text{Li,0}DF}{j_\text{SEI,0,0}}e^{-(1-\alpha_\text{SEI})\tilde{\eta}_\text{SEI}} ,\\
\label{eq:L_mig}
L_\text{mig}&=\frac{2RT\kappa_{\text{Li}^+,\text{SEI}}}{F \left|j_\text{int}\right|} .
\end{align}
For realistic parameters, $L_\text{diff} \ll L_\text{mig}$ holds (see supporting information (SI), table \ref{tab:model_parameters}).

We assume that each electron reaching the SEI-electrolyte interface is instantly consumed by SEI formation. Thus, we link the SEI current $j_\text{SEI}$ directly to the SEI growth rate $\text{d}L_\text{SEI}/\text{d}t$, 
\begin{equation}
\label{eq:relationship_charge_thickness}
\frac{dL_\text{SEI}}{dt}=-\frac{V_\text{SEI}}{F} j_\text{SEI}
\end{equation}
with the mean molar volume of SEI components $V_\text{SEI}$. 
Based on equation \ref{eq:relationship_charge_thickness}, we proceed analyzing the growth behavior of the SEI with respect to the elapsed time $t$. To this aim, we insert the SEI current $j_\text{SEI}$ (see equation \ref{eq:SEI_current_final}), into the growth rate $\text{d}L_\text{SEI}/\text{d}t$ (see equation \ref{eq:relationship_charge_thickness}). 
In the following, we derive analytic solutions of the resulting differential equation for three different limiting cases. We compare them with the full numeric solution in sections \ref{s:Results} and \ref{s:Discussion}. 

\textit{First}, if the SEI is thin, i.e. $L_\text{app}\ll L_\text{diff}$, we can simplify the SEI current to equation \ref{eq:Reaction_limitation},
\begin{equation}
\label{eq:Reaction_limitation}
j_{\text{SEI,re}}=-j_{\text{SEI,0,0}}e^{-\alpha_\text{SEI}\tilde{\eta}_\text{SEI}}.
\end{equation}
Thus, in this regime, SEI growth is limited by the formation reaction of neutral lithium atoms. Inserting equation \ref{eq:Reaction_limitation} into the SEI growth equation \ref{eq:relationship_charge_thickness} yields a linear SEI growth in time,
\begin{align}
L_\text{SEI} &=\frac{V_\text{SEI}}{F} j_{\text{SEI,0,0}}e^{-\alpha_\text{SEI}\tilde{\eta}_\text{SEI}} \cdot t .
\end{align}

\textit{Second}, if $L_\text{diff} \ll L_\text{app} \ll L_\text{mig}$, we get 
\begin{equation}
\label{eq:Transport_limitation}
j_{\text{SEI,diff}}=-\frac{c_\text{Li,0}DF}{L_\text{SEI}}	e^{-\tilde{\eta}_\text{SEI}}.
\end{equation}
Here, diffusion of lithium interstitials limits SEI growth, which results in a SEI growth proportional to $\sqrt{t}$,
\begin{align}
\label{eq:SEI_growth_Tr}
\begin{split}
L_\text{SEI} &=L_\text{tun} \\
&+\sqrt{2V_\text{SEI}c_\text{Li,0}De^{-\tilde{\eta}_\text{SEI}} \cdot t +\left(L_\text{SEI,0}-L_\text{tun}\right)^2}.
\end{split}
\end{align}
This form of SEI current and growth coincides with the form derived by Single et al. \cite{Single2018} in the case of battery storage, i.e. $\eta_\text{int}=0$. For battery operation, the intercalation overpotential $\eta_\text{int}$ affects $\tilde{\eta}_\text{SEI}$ according to equation \ref{eq:Overpotential_SEI_2} and thus accelerates SEI growth during charging and decelerates SEI growth during discharging. 

\textit{Third}, if $L_\text{mig} \ll L_\text{app}$, the SEI current has the form shown in equations \ref{eq:Migration_limitation_charge} and \ref{eq:Migration_limitation_discharge},
\begin{subequations}
\label{eq:Migration_limitation}
\begin{align}
&j_{\text{SEI,mig}}&&=\frac{c_\text{Li,0}DF^2j_\text{int}}{2RT\kappa_{\text{Li}^+,\text{SEI}}}e^{-\tilde{\eta}_\text{SEI}} &&\text{charging}, \label{eq:Migration_limitation_charge}\\
&j_{\text{SEI,mig}}&&=0 &&\text{discharging}. \label{eq:Migration_limitation_discharge}
\end{align} 
\end{subequations}
In this regime, migration of electrons through the SEI becomes dominant.
SEI formation is irreversible, so that the SEI current must be negative. Thus, we have to distinguish between charging and discharging in this case. While SEI growth is fully suppressed during discharging, equation \ref{eq:SEI_growth_Mig} describes growth during charging.
\begin{equation}
\label{eq:SEI_growth_Mig}
L_\text{SEI} =\frac{V_\text{SEI}c_\text{Li,0}DFj_\text{int}}{2RT\kappa_{\text{Li}^+,\text{SEI}}}e^{-\tilde{\eta}_\text{SEI}} \cdot t 
\end{equation}

\section{Numerical Methods}
\label{s:Methods}
We briefly summarize the implementation of our model developed in the previous section before we simulate SEI growth during battery cycling in the following section. We model galvanostatic battery operation and thus apply a constant current $j$, which leads to the intercalation current $j_\text{int}=j-j_\text{SEI}$ according to equation \ref{eq:j_ges}. The intercalation current $j_\text{int}$ affects the lithium concentration inside the anode $c_\text{s}$ according to the differential equation \ref{eq:DAE1}. Thereby, also the OCV $U_0$ changes according to the $U_0(c_\text{s})$-curve measured by Attia et al. \cite{Attia2019} (see \ref{eq:U_of_Q}).
Growth of SEI thickness is described by equation \ref{eq:relationship_charge_thickness} with the SEI current $j_\text{SEI}$ from equation \ref{eq:SEI_current_final}. In order to calculate the apparent thickness $L_\text{app}$, we use a continuous function, which smooths the transition between the tunneling and the diffusion-migration regime (see equation \ref{eq:apparent_thickness}). 
Equation \ref{eq:DAE1}, equation \ref{eq:relationship_charge_thickness} and the galvanostatic condition give a differential algebraic system of equations (DAE), which simultaneously describes battery operation and SEI growth.

We iteratively solve this DAE along the elapsed time with the ordinary differential equation solver ode15s of MATLAB. The simulation stops, when it reaches the end-of-charge voltage $U_1$ or the end-of-discharge voltage $U_2$. We transform the current densities, given in C-rate, to $\si{\ampere\per\square\meter}$ with equation \ref{eq:transform_C_Am2}
\begin{equation}
\label{eq:transform_C_Am2}
j\left[\si{\ampere\per\square\meter}\right]=\frac{Q_\text{s,nom}}{\SI{1}{\hour}}\cdot \frac{1}{A_\text{cb}} \cdot j\left[\text{C-rate}\right],
\end{equation}
using the nominal capacity $Q_\text{s,nom}$.
Table \ref{tab:model_parameters} in the supporting information lists the parameter of the model.

Based on the results of the DAE, we simulate the differential capacity analysis experiments of Attia et al. \cite{Attia2019} with equation \ref{eq:dQ_dU_sim},
\begin{equation}
\label{eq:dQ_dU_sim}
\frac{\text{d}Q_\text{SEI}}{\text{d}U_0}_\text{sim}=\frac{\text{d}Q}{\text{d}U_0}_\text{sim}-\frac{\text{d}Q}{\text{d}U_0}_\text{baseline}
\end{equation}
with the simulated differential capacity $\frac{\text{d}Q}{\text{d}U_0}_\text{sim}=j \frac{\text{d}t}{\text{d}U_0}$.
We calculate the baseline differential capacity $\frac{\text{d}Q}{\text{d}U_0}_\text{baseline}$ from the open-circuit voltage $U_0(c_\text{s})$ (see equation \ref{eq:dQ_dU_baseline}).
The SEI capacity per cycle $n$, $Q_\text{SEI}(n)$, is obtained from integration of equation \ref{eq:dQ_dU_sim} over the voltage region,
\begin{equation}
\label{eq:Q_per_cycle}
Q_\text{SEI}(n)=\int_{U_1}^{U_2}\frac{\text{d}Q}{\text{d}U_0}_\text{sim}(n)\text{d}U_0 .
\end{equation}
The overall charge consumption $Q(n)$ results from equation \ref{eq:Q_per_cycle} by adding a constant intercalation capacity $Q_\text{s}(j)$.

\section{Results}
\label{s:Results}
In the following, we compare our theory described in section \ref{s:Theory} with the experiments of Attia et al. \cite{Attia2019} on different time-scales. First, we investigate the voltage and current dependence of the short-term SEI growth in section \ref{ss:Short}. Second, we analyze the temporal evolution of SEI growth in the long-term in section \ref{ss:Long} ($2<n<1000$). Third, we investigate the time dependence of SEI growth for very long times in section \ref{ss:Ultra_Long} ($100<n$).

\subsection{Short-term SEI growth}
\label{ss:Short}
\begin{figure}[tb] 
 \centering
 \includegraphics[width=8.4 cm]{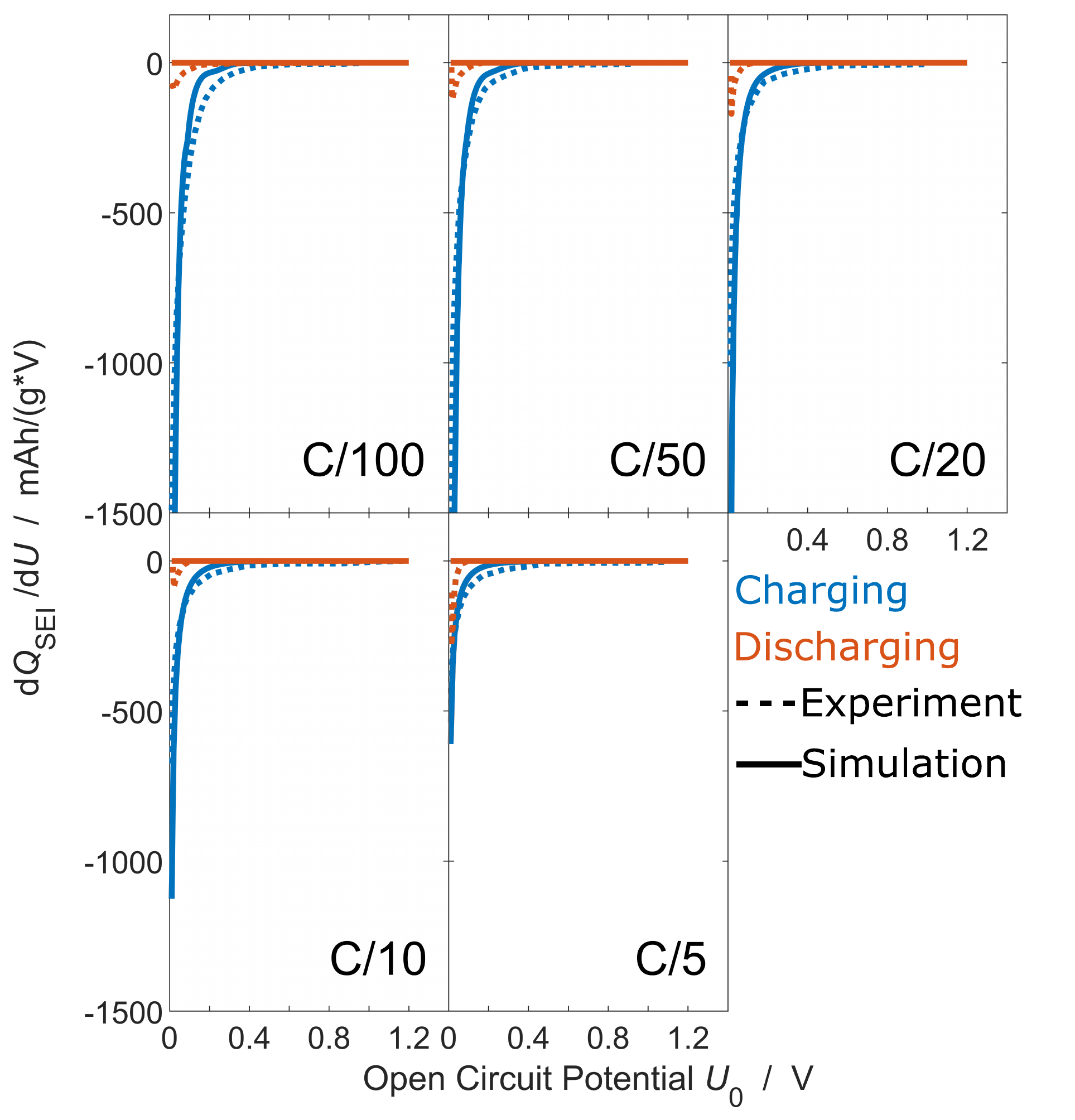}
 \caption{Consumed SEI capacity during the second cycle as function of OCV for different applied currents $j=\text{C}/100,\text{C}/50,\text{C}/20,\text{C}/10,\text{C}/5$. We compare experiments \cite{Attia2019} (dashed) and simulation results (solid, equation \ref{eq:dQ_dU_sim}). Charging is depicted in blue, discharging in orange.}
 \label{fig:dQ_dU}
\end{figure}

We compare the differential capacity analysis experiments $\text{d}Q_\text{SEI}/\text{d}U_0$ of Attia et al. \cite{Attia2019} with the results of our simulation in figure \ref{fig:dQ_dU}.
The $\text{d}Q_\text{SEI}/\text{d}U_0$ curve depends exponentially on the cell voltage. Our simulations agree with this behavior for all charging currents. For discharging currents, however, we observe a deviation between experiments and simulations.

A reaction kinetic limitation causes this exponential voltage dependence. We rationalize this behavior with the approximation of the SEI current $j_\text{SEI}$ for thin layers in equation \ref{eq:Reaction_limitation}. Inserting the definition of the $\ce{Li}^0$ formation potential jump $\tilde{\eta}_\text{SEI}$ (see equation \ref{eq:Overpotential_SEI_2}) leads to 
\begin{equation}
\label{eq:Short_term_explanation}
j_{\text{SEI}}=-j_{\text{SEI,0,0}}e^{-\alpha_\text{SEI}\frac{F}{RT}(\eta_\text{int}+U_0+\mu_\text{Li,0}/F)}.
\end{equation}
Thus, the SEI current $j_\text{SEI}$ depends exponentially on the OCV $U_0$. The exponential factor $\alpha_\text{SEI}=0.22$ agrees with the experimentally determined one.

\begin{figure}[tb] 
 \centering
 \includegraphics[width=8.4 cm]{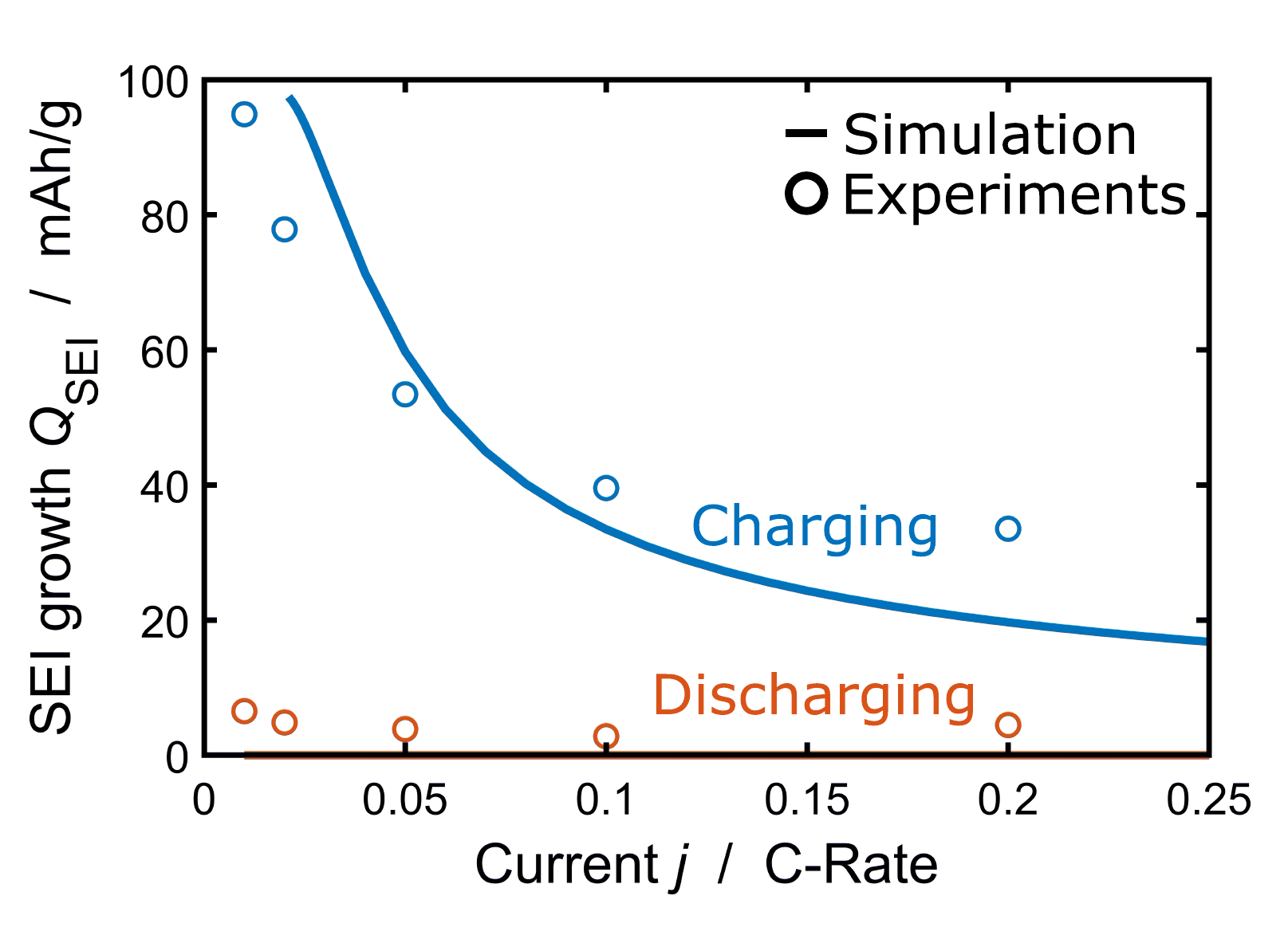}
 \caption{Current dependence of the overall SEI charge during the second cycle. We compare experiments \cite{Attia2019} (circles) and simulation results (line, equation \ref{eq:Q_per_cycle}). Charging is depicted in blue, discharging in orange.}
 \label{fig:Q_vs_j}
\end{figure}

The asymmetry factor $\alpha_\text{SEI}$ is indispensable for modeling the experimentally observed voltage dependence in figure \ref{fig:Q_vs_j}. This proofs that reaction kinetics govern the second-cycle SEI growth. In contrast, long-term growth models \cite{Single2018} assume equilibrium at the electrode surface and are governed by the growth law in equation \ref{eq:Transport_limitation}. This growth law lacks the asymmetry factor $\alpha_\text{SEI}$ and thus deviates from the experiments of Attia et al. \cite{Attia2019} We conclude that second-cycle SEI growth cannot be explained with equilibrium reaction conditions, but it can be explained with appropriate reaction kinetics.

The value $\alpha_\text{SEI}=0.22$ \cite{Li2019,Crowther2008} points to complex reaction kinetics consisting of different phenomena, which we do not resolve in our lumped Butler-Volmer kinetics in equations \ref{eq:Butler-Volmer} and \ref{eq:Butler_Volmer_SEI}. For example, change of electron bands at the interfaces, enhanced electron tunneling, and capacitive effects may play a role. Interestingly, in the low voltage regime, the OCV-curve measured by Attia et al. \cite{Attia2019} (see equation \ref{eq:U_of_Q}) shows the same exponential behavior as the SEI formation current \ref{eq:Short_term_explanation}. This indicates that unresolved surface processes occur.

During discharge, experiments and simulations disagree. We attribute this to a retardation effect. The experiments of Attia et al. \cite{Attia2019} immediately switch from charging to discharging. Thus, capacitive processes originating from the end of charging affect the discharging. Our model, however, does not resolve such capacitive processes like the lithium ion concentration throughout the SEI.
Das et al. \cite{Das2019} have modeled the experiments of Attia et al. \cite{Attia2019}. Their equations describe the same ideal diode effect during discharging which should also suppress SEI growth during discharging. Furthermore, the modeling approach of Das et al. \cite{Das2019} exhibits large overpotentials due to concentration polarization. In our simulations, we observe these high intercalation overpotentials, too.

Next, we analyze the influence of the curent $j$ on the total SEI growth $Q_\text{SEI}$ in the second cycle. We determine $Q_\text{SEI}$ by equation \ref{eq:Q_per_cycle} and compare it to the experiments of Attia et al. \cite{Attia2019} in figure \ref{fig:Q_vs_j}. 
Our simulation results follow the experimentally measured trends. We observe a strong asymmetry between charging and discharging. During discharging, second cycle SEI growth is suppressed. Charging, in contrast, enhances SEI growth and $Q_\text{SEI}$ increases with decreasing current. 

Two opposing trends determine the current dependence of SEI growth per cycle during charging. On the one hand, SEI growth per cycle decreases with increasing current, because the cycle time decreases according to $t_\text{cycle}=Q_\text{s,max}/j_\text{int}$. On the other hand, SEI growth increases with increasing current due to the intercalation overpotential $\eta_\text{int}$ (see equation \ref{eq:Short_term_explanation}).
Let us calculate the dependence of $j_\text{SEI}$ on $j_\text{int}$. The SEI current $j_\text{SEI}$ in equation \ref{eq:Short_term_explanation} depends on $\eta_\text{int}$. We determine $\eta_\text{int}$ in terms of $j_\text{int}$ by inverting equation \ref{eq:Butler-Volmer} in the Tafel regime (see equation \ref{eq:Tafel_equation_2}). Combining both contributions, the second-cycle SEI capacity $Q_\text{SEI}$ scales with the intercalation current $j_\text{int}$ according to
\begin{equation}
\label{eq:Short_term_current}
Q_\text{SEI} \propto \frac{\left(j_\text{int}\right)^{2\alpha_\text{SEI}}}{j_\text{int}}.
%\hspace{10 mm}\text{Reaction limitation}.
\end{equation}
%&&
%\label{eq:Short_term_current_diffusion}
%&Q_\text{SEI} &&\propto \frac{\left(j_\text{int}\right)^2}{j_\text{int}} &&\text{Diffusion limitation}
We analyze the implications of the asymmetry factor $\alpha_\text{SEI}$ on the observed current dependence depicted in figure \ref{fig:Q_vs_j} based on equation \ref{eq:Short_term_current}. For our choice $\alpha_\text{SEI}=0.22$, we obtain a decreasing $Q_\text{SEI}(j_\text{int})$ in agreement with the experiments. We note that $\alpha_\text{SEI}=0.31$ would give the best agreement of our simulations with experiments with respect to the current dependence. Attia and Das et al. \cite{Attia2019,Das2019} disuss the current dependence by plotting $Q_\text{SEI}/t_\text{cycle}$ versus $j_\text{int}$. They conclude that $Q_\text{SEI}/t_\text{cycle}$ is linear in $j_\text{int}$, i.e., that $Q_\text{SEI}$ is independent of current. This disagrees with their experimental data reprinted in figure \ref{fig:Q_vs_j}.

At small applied currents $j<0.05\text{C}$, the entanglement of intercalation current and SEI current in the applied current $j=j_\text{int}+j_\text{SEI}$ constitutes a fundamental challenge for modeling. Therefore, we do not plot simulation results for small current in figure \ref{fig:Q_vs_j}. In this case, the SEI current $j_\text{SEI}$ becomes twice as large as the intercalation current $j_\text{int}$. The suppressed intercalation current $j_\text{int}$ leads to a long cycle time and a large SEI capacity $Q_\text{SEI}$. Thus, at small currents, the SEI thickness crosses the critical diffusion thickness ($L_\text{app}>L_\text{diff}$) during the second cycle and diffusion dominates SEI growth (see equation \ref{eq:Transport_limitation}). This leads to an increasing course of $Q_\text{SEI} \propto {\left(j_\text{int}\right)^2}/{j_\text{int}}$. To sum up, for $j\lesssim 0.05\text{C}$, our simulation results deviates from the scaling law in equation \ref{eq:Short_term_current}. 

This deviation results from our method of electron counting. Our model relies on the idea that electron consumption for SEI growth and intercalation occur simultaneously. This assumption leads to the reaction equations \ref{eq:SEI_intercalation_reaction} and \ref{eq:SEI_interstitial_reaction} as well as the relationship $j=j_\text{int}+j_\text{SEI}$. In reality, however, $\ce{Li^0}$ can also result indirectly from intercalated lithium $\ce{Li_xC_6}$ according to reaction equation \ref{eq:SEI_interstitial_reaction_2},
\begin{equation}
\label{eq:SEI_interstitial_reaction_2}
\ce{Li_xC_6} \rightleftharpoons x\ce{Li^0} + \ce{C_6}.
\end{equation}
In this approach, the intercalation current would equal the applied current $j=j_\text{int}$, so that intercalation would not be suppressed even for low $j$. However, in order to keep our model as simple as possible, we neglect this option for $\ce{Li^0}$ formation.

\subsection{Long-term SEI growth}
\label{ss:Long}
\begin{figure}[tb] 
 \centering
 \includegraphics[width=8.4 cm]{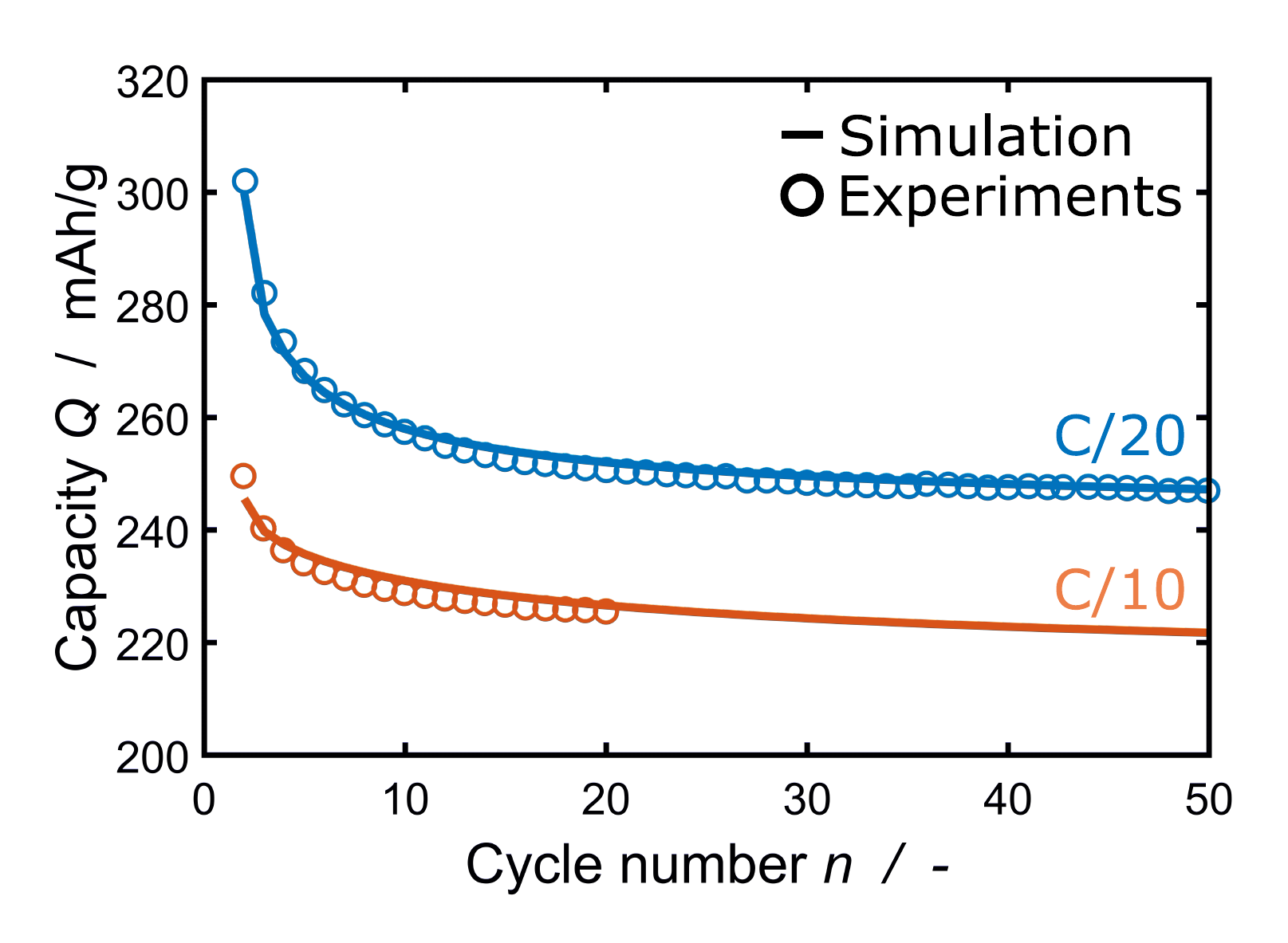}
 \caption{Development of the overall charge consumed for SEI formation over several cycles. We compare experiments \cite{Attia2019} (circles) and simulation results (solid lines, equation \ref{eq:Q_per_cycle}). C/20 in blue, C/10 in orange.}
 \label{fig:Q_vs_n}
\end{figure}

We continue to analyze the SEI capacity $Q_\text{SEI}$ and how it evolves with increasing cycle number $n$. Figure \ref{fig:Q_vs_n} compares the simulation results for the overall capacity $Q(n)$ determined by equation \ref{eq:Q_per_cycle} with the experiments of Attia et al. \cite{Attia2019}. 
We observe that the consumed capacity decreases with each cycle and that the simulation nicely fits the experiment. Comparing the different applied currents, we notice that $Q_\text{SEI}$ decreases faster for C/20 compared to C/10.

The observed decrease in SEI capacity $Q_\text{SEI}$ per cycle $n$ stems from transport limited SEI growth. In this regime, our model agrees to the model for neutral lithium diffusion of Single et al. \cite{Single2018} Thus, in contrast to the model of Das et al. \cite{Das2019}, our model predicts the well-known $\sqrt{t}$ time dependence of the overall SEI growth $L_\text{SEI}$ (see equation \ref{eq:SEI_growth_Tr}), for long times. 

Based on the growth law in this limit (see equation \ref{eq:SEI_growth_Tr}), we derive the dependence of SEI growth $Q_\text{SEI}(n)$ on cycle number $n$. To this aim, we link the cycle number $n$ to the elapsed time $t=Q_\text{s,max}/j_\text{int}\cdot n$ and the overall SEI charge consumption to the SEI thickness $L_\text{SEI}=V_\text{SEI}/F\cdot Q_\text{SEI,tot}$. Taking the derivative of $L_\text{SEI}$ with respect to $n$ (see equation \ref{eq:SEI_growth_Tr}) yields the capacity fade per cycle $Q_\text{SEI}(n)$,
\begin{align}
\label{eq:Q_n_dependence}
Q_\text{SEI}(n)=&\frac{\text{d}Q_\text{SEI,tot}}{\text{d}n}\\
=&\left[\frac{V_\text{SEI}}{2 c_\text{Li,0}DF^2}\cdot \frac{j_\text{int}}{e^{-\tilde{\eta}_\text{SEI}}} \cdot n \right. \nonumber\\
&\hspace{15 mm}\left. + \left(\frac{F(L_\text{SEI,0}-L_\text{tun})}{V_\text{SEI}} \right)^2\right]^{-1/2} .\nonumber
\end{align}
Thus, $Q_\text{SEI}(n)$ decays monotonously with the inverse of the cycle number as $1/\sqrt{n}$.
The slope depends on the current in the form $j_\text{int}/e^{-F\eta_\text{int}/RT} \approx 1/j_\text{int}$ (see equation \ref{eq:Overpotential_SEI_2} and equation \ref{eq:Tafel_equation_2}) and is thus larger for C/20 than for C/10.

\subsection{Ultra long-term SEI growth}
\label{ss:Ultra_Long}
We proceed by analyzing the SEI growth for very long times ($100<n$).
\begin{figure}[tb] 
 \centering
 \includegraphics[width=8.4 cm]{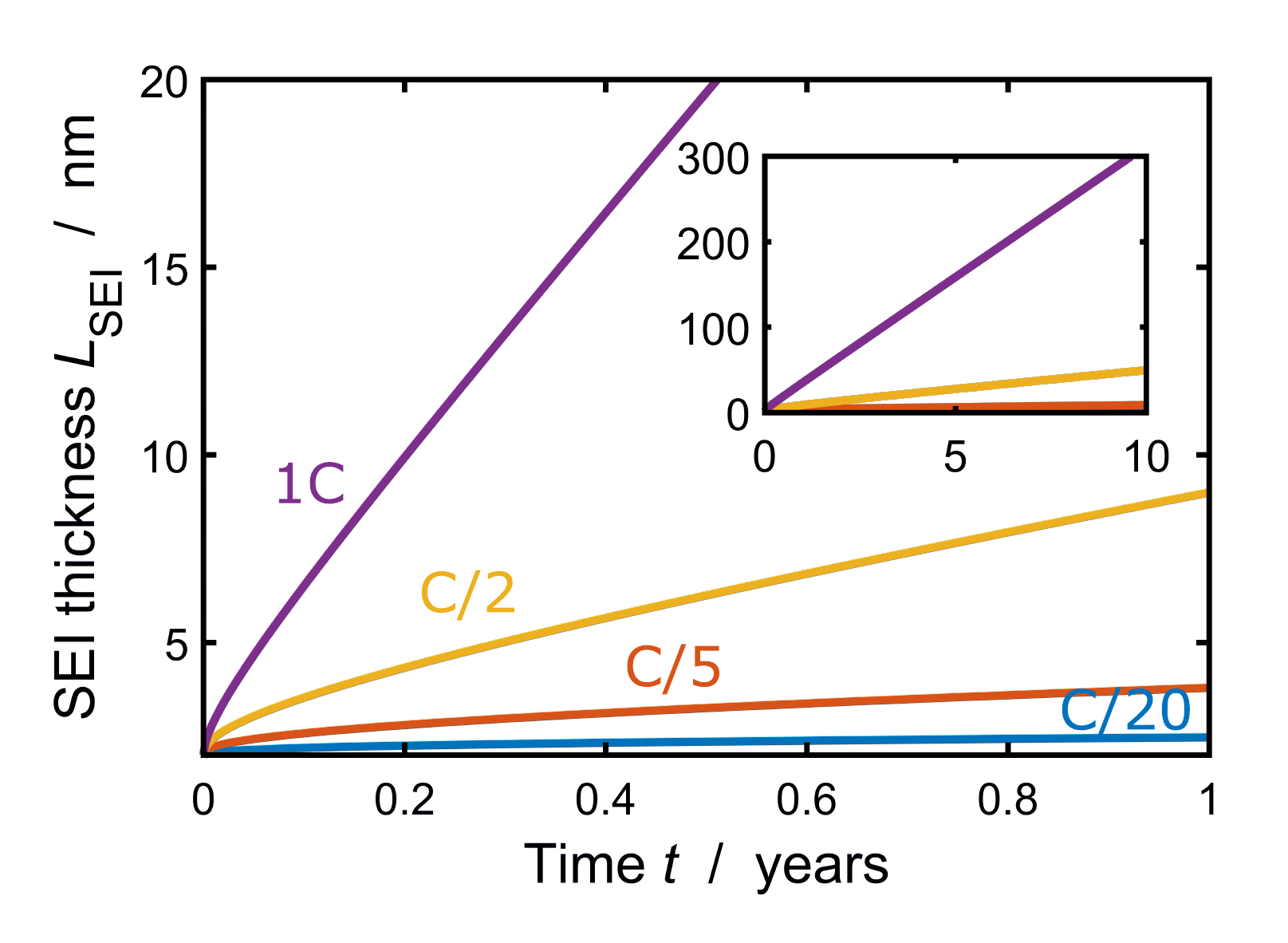}
 \caption{SEI thickness $L_\text{SEI}$ with respect to time $t$ for continuous cycling of graphite in a SoC range of $0.2 \leq c_\text{s}/c_\text{s,max}\leq 0.8$ for different applied currents $j=C/20,C/5,C/2,1C$.}
 \label{fig:Thickness_Mig}
\end{figure}
In figure \ref{fig:Thickness_Mig}, we show the growth of SEI thickness $L_\text{SEI}$ over time $t$ for continuous cycling of a graphite anode at various currents $j$ (see table \ref{tab:model_parameters_graphite})\cite{Shornikova2009,Tarascon2010,Kipling1964a}. We observe that the SEI thickness grows faster for higher charging currents. Additionally, the slope of the curves changes over time, starting from a square-root-of-time-dependence and shifting towards a linear time-dependence.

SEI growth is faster for higher currents, because the SEI current increases with the intercalation current $j_\text{int}$ according to equations \ref{eq:Transport_limitation} and \ref{eq:Migration_limitation}. The cause for the transition in time dependence is a shift from diffusion limited to migration limited growth. Over time, the SEI thickness $L_\text{SEI}$ grows and approaches the critical migration thickness $L_\text{mig}$. Below the transition thickness, diffusion limits SEI growth according to equation \ref{eq:SEI_growth_Tr} leading to a $\sqrt{t}$-time dependence. Above the transition thickness, electromigration is the growth limiting process, which results in a $t$-time dependence of the curve, according to equation \ref{eq:SEI_growth_Mig}.

A shift to linear SEI growth was so far observed by different experimental groups \cite{Ekstrom2015,Li2015,Groot2015,Keil2019,Yoon2020}. This transition is typically attributed to mechanical effects e.g. repeated SEI fracture and regrowth \cite{Pinson2012,Li2015,Perassi2019}. Our approach shows a complementary explanation of linear SEI growth within electrochemistry.

\section{Discussion}
\label{s:Discussion}
In the previous section, we reveal that different growth mechanisms are dominant at different time scales. We follow this line of thought in this section and systematically analyze the transition between the growth regimes. We first calculate the SEI current magnitude depending on the operating conditions and study the asymmetry between charging and discharging in section \ref{ss:SEI_current}. Subsequently, we analyze the influence of operating conditions on the transition between the regimes in section \ref{ss:Transition}. First, we investigate the transition from reaction to diffusion limitation. Second, we look at the transition between diffusion and electromigration limitation.

\subsection{Asymmetry between charging and discharging}
\label{ss:SEI_current}

We analyze how the operating conditions influence the SEI growth rate $\text{d}L_\text{SEI}/\text{d}t$. To this aim, we take a look at the growth rate for various currents $j$ and OCVs $U_0$ with an SEI thickness of $L_\text{SEI}=\SI{3}{\nano\meter}$.
\begin{figure}[tb] 
 \centering
\includegraphics[width=8.4 cm]{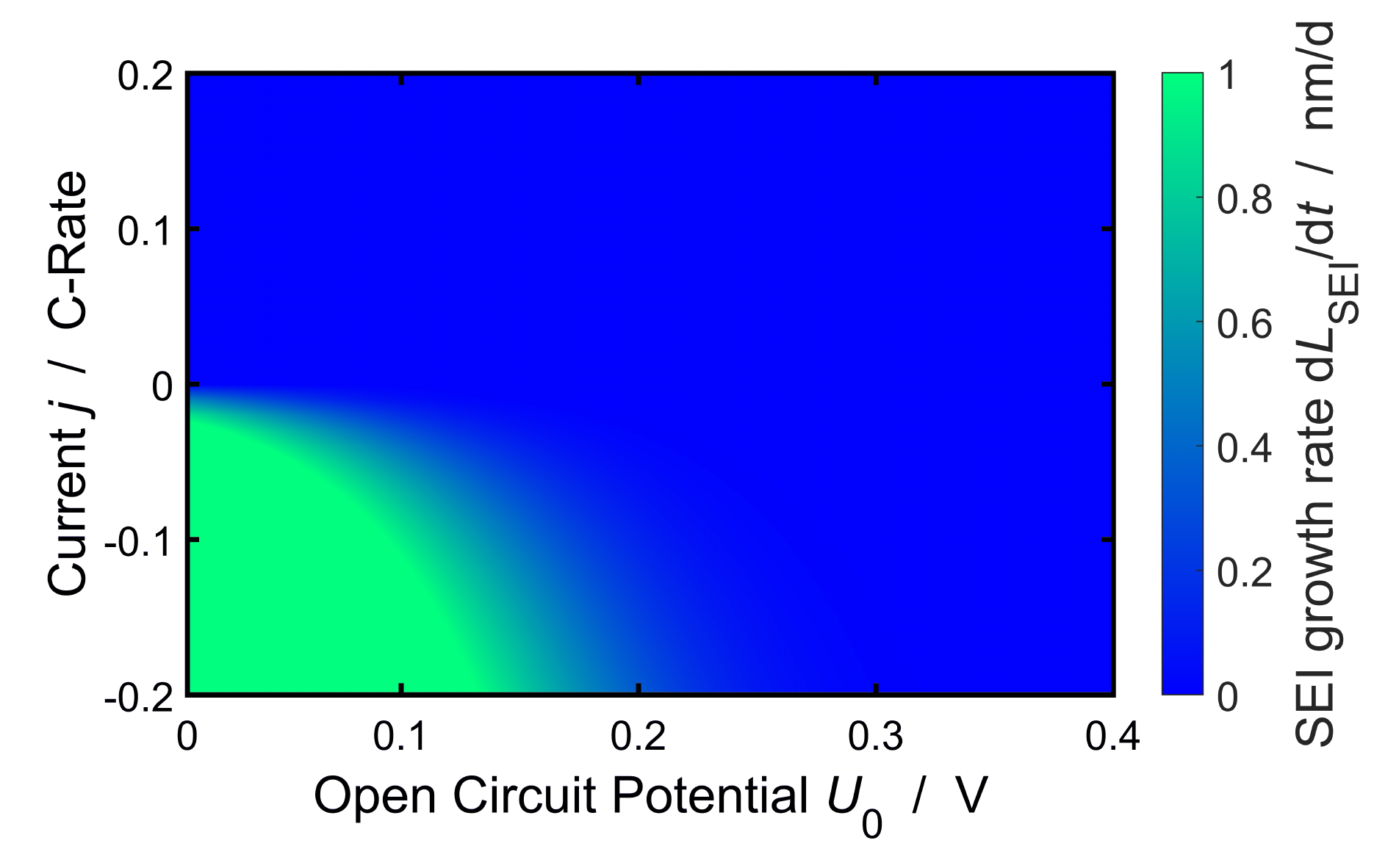}
 \caption{SEI growth rate with respect to applied current and open circuit potential for an SEI tickness of $L_\text{SEI}=\SI{3}{\nano\meter}$ (see equation \ref{eq:relationship_charge_thickness}).}
 \label{fig:SEI_magnitude}
\end{figure}
Figure \ref{fig:SEI_magnitude} clearly shows the asymmetry between charging and discharging: SEI grows fast during charging and slow during discharging. Furthermore, low electrode voltages accelerate SEI growth. Both trends result from the SEI overpotential $\tilde{\eta}_\text{SEI}$ (equation \ref{eq:Overpotential_SEI_2}), which exponentially increases the SEI current for low voltages and high intercalation currents. These results show that the capacity of lithium-ion batteries fades fastest for high state-of-charge and high charging rate.

\subsection{Transition between regimes}
\label{ss:Transition}
\begin{figure*}[htb] 
 \centering
\includegraphics[width=17.8 cm]{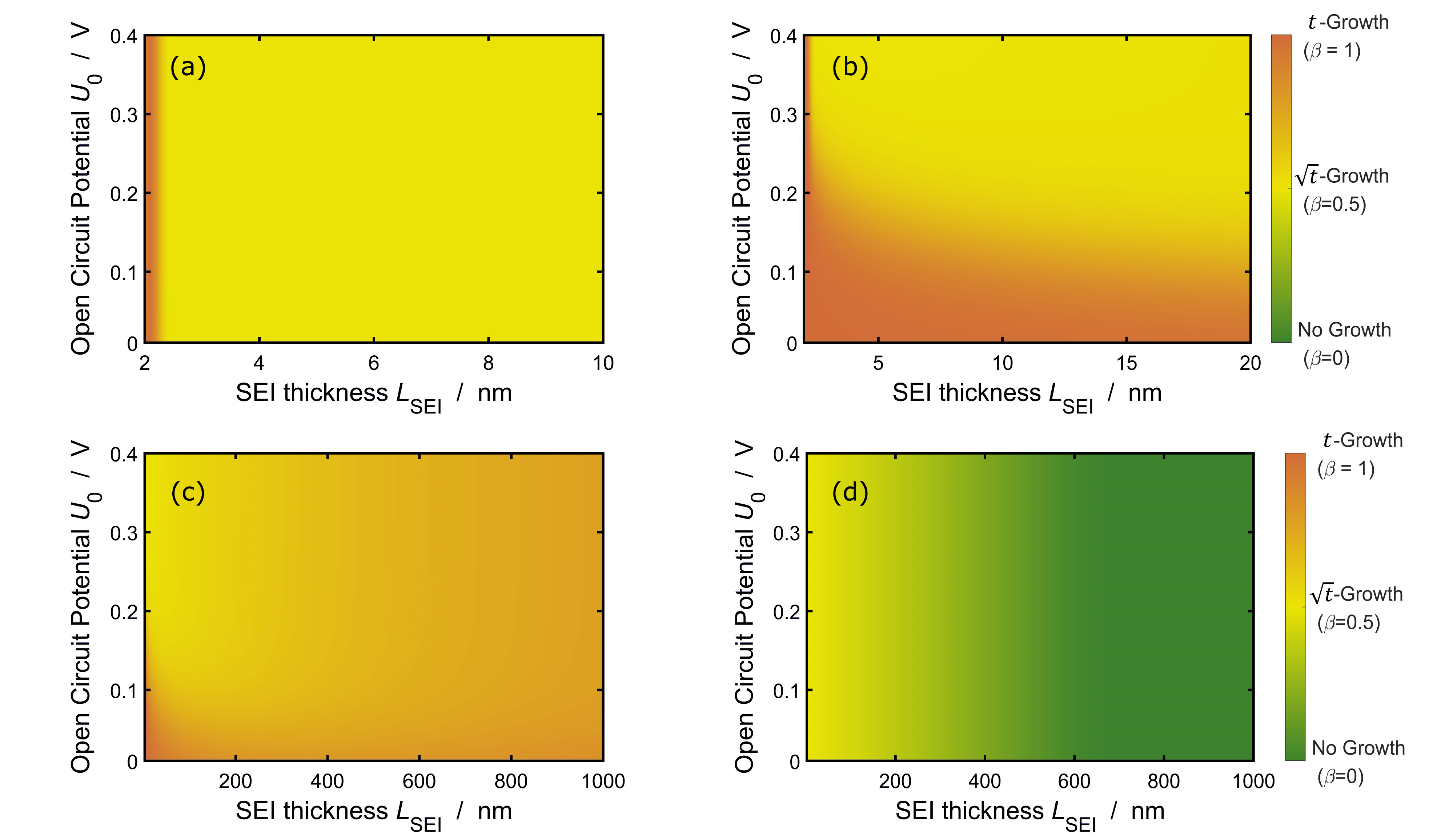}
 \caption{Scaling factor $\beta$ (see equation \ref{eq:Time_dependence}) of time dependence of SEI growth as a function of open circuit voltage and SEI thickness according to equation \ref{eq:relationship_charge_thickness}. Red indicates reaction limitation (equation \ref{eq:Reaction_limitation}, $\beta=1$) or migration limitation during charging (equation \ref{eq:Migration_limitation_charge}, $\beta=1$). Yellow indicates diffusion limitation (equation \ref{eq:Transport_limitation}, $\beta=0.5$) and green migration limitation (equation \ref{eq:Migration_limitation_discharge}, $\beta=0$). (a) Battery storage. (b) Battery charging with $j=-0.2\text{C}$ in the short-term. (c) Battery charging with $j=-0.2\text{C}$ in the long-term. (d) Battery discharging with $j=0.2\text{C}$ in the long-term.}
 \label{fig:Regimes_short}
\end{figure*}
We proceed by identifying the different dominant growth mechanisms based on the respective time dependence of SEI growth, $L_\text{SEI}(t)$. To this aim, we express the scaling of SEI thickness with time in the general form shown in equation \ref{eq:Time_dependence},
\begin{equation}
\label{eq:Time_dependence}
L_\text{SEI}\propto t^\beta \Leftrightarrow \beta=\frac{\text{d}\log (L_\text{SEI})}{\text{d}\log (t)}.
\end{equation}
The parameter $\beta$ indicates the dominant growth mechanism according to 
\begin{itemize}
\item $\beta=1$: reaction limitation or \\    migration limitation during charging,
\item $\beta=0.5$: diffusion limitation,
\item $\beta=0$: migration limitation during discharging.
\end{itemize}
$\beta$ depends on the applied current $j$, the OCV $U_0$ and the SEI thickness $L_\text{SEI}$.
First, we look at the growth behavior during storage in figure \ref{fig:Regimes_short}a. We observe a sharp transition between reaction and diffusion limitation for the SEI thickness $L_\text{SEI}\approx\SI{2.4}{\nano\meter}$, which is independent of the open-circuit potential $U_0$. The tunneling thickness $L_\text{tun}$ is the reason for this transition. Below this thickness, electrons easily tunnel through the SEI, so that the SEI formation is limited by the $\ce{Li^0}$ reaction kinetics. Above this thickness, diffusion through the SEI becomes dominant leading to a transport limitation in agreement with the measurements of Keil et al. \cite{Keil2019} and the model of Single et al. \cite{Single2018}.

During battery charging (see figure \ref{fig:Regimes_short}b), the transition between reaction and diffusion limitation is smeared out. We observe in figure \ref{fig:Regimes_short}b that SEI growth is reaction limited for a thin SEI and a low OCV $U_0$. Diffusion limits growth for a high OCV $U_0$ and a thick SEI.

To understand this behavior, we recall the premise for reaction limitation derived in the theory section, $L_\text{app} \ll L_\text{diff}$. This is fulfilled for low SEI thicknesses $L_\text{app}$ or large critical diffusion thicknesses $L_\text{diff}$. According to equation \ref{eq:L_diff}, the critical diffusion thickness $L_\text{diff}$ grows exponentially with decreasing $\ce{Li^0}$ formation potential jump $\tilde{\eta}_\text{SEI}$ and thereby with decreasing $U_0$ (see equation \ref{eq:Overpotential_SEI_2}). We thus observe reaction limitation for low OCV $U_0$ and low SEI thicknesses $L_\text{SEI}$.

The transition from reaction to diffusion limitation has important implications for the current-, OCV- and time-dependence of SEI growth (see equations \ref{eq:Reaction_limitation} and \ref{eq:Transport_limitation}). For reaction limited SEI growth, the SEI thickness scales with $t$; for diffusion limited SEI growth, it scales with $\sqrt{t}$. OCV- and current dependence are weaker for reaction limitation due to the exponential factor $\alpha_\text{SEI}$. Reaction limitation exhibits an exponential dependence on the OCV, weakened by $\alpha_\text{SEI}$, and a sub-linear dependence on the current $j$. In contrast, transport limitation shows an exponential dependence on the OCV and a quadratic current dependence.

Next, we analyze the growth behavior of the SEI for longer times in figures \ref{fig:Regimes_short}c and d.
We observe a continuous transition from transport (yellow) to migration (red in \ref{fig:Regimes_short}c, green in \ref{fig:Regimes_short}d) limitation for all voltages. 

This transition arises as the SEI thickness $L_\text{app}$ approaches the critical migration thickness $L_\text{mig}$, defined by equation \ref{eq:L_mig}. This shift in limiting mechanism leads to a shift in the time dependence of SEI thickness from $\sqrt{t}$ to $t$ (during charging) respective constant (during discharging) according to equations \ref{eq:Transport_limitation} and \ref{eq:Migration_limitation}. We note that the current dependence is stronger for migration limitation.

Summarizing figures \ref{fig:Regimes_short}b, c and d, we observe a transition in the time dependence of SEI growth from $t\rightarrow\sqrt{t}\rightarrow t / (\text{const.})$ due to a shift in the dominant formation mechanism from reaction to diffusion to migration limited. This finding explains phenomenologically derived capacity fade equations of the form 
\begin{equation}
\label{eq:Phenomenlogical}
\Delta Q\propto t^\beta \qquad 0 \leq \beta \leq 1,
\end{equation}
as transition between either diffusion and reaction or diffusion and migration limitation \cite{Kabitz2013,Schmalstieg2014}. Moreover, our findings show that linear capacity fade is inherent to the electrochemistry of the system and not necessarily caused by SEI fracture and reformation \cite{Pinson2012,Li2015,Perassi2019,Yoon2020}.

\section{Conclusion}
\label{s:Conclusion}

In summary, we have extended an existing model for SEI growth during battery storage \cite{Single2018} to incorporate the effects of battery operation. A comparison of the model predictions with the experiments of Attia et al. \cite{Attia2019} showed very good agreement. Based on the so-validated model we proceed analyzing the SEI growth behavior in detail. We find that the formation reaction of neutral lithium atoms initially limits SEI growth. With increasing SEI thickness, first diffusion and then electromigration of the electrons coordinated to lithium ions limits further SEI growth. The resulting model for diffusion limitation agrees with the model of Single et al. \cite{Single2018} in the case of battery storage.

Our novel modeling approach predicts a shift in time dependence of capacity fade from $t\rightarrow\sqrt{t}\rightarrow t/\text{const.}$ over time. For the first time, the time dependence explains the so far empirically motivated capacity fade equations of the form $\Delta Q \propto t^\beta$ with $0 \leq \beta \leq 1$ as transitions between transport and reaction limited growth \cite{Kabitz2013,Schmalstieg2014,Christensen2004}. Moreover, these new insights show that besides SEI fracture and reformation the inherent electrochemistry of SEI growth leads to a linear SEI growth during long-term battery cycling.

Our theory can be extended to account for lithium plating, i.e., the precipitation of lithium atoms $\ce{Li^0}$ at the anode, as we model $\ce{Li^0}$ as mediator for SEI growth. The amount of $\ce{Li^0}$ in the SEI exponentially increases at low potentials, when lithium plating occurs. In order to resolve inhomogeneous SEI growth and lithium plating caused by locally varying operating conditions, the theory developed in this work can be implemented into a three dimensional battery simulation.

\section*{Acknowledgments}

We gratefully acknowledge funding and support by the German Research Foundation (DFG) within the research training group SiMET under the project number 281041241/GRK2218. The support of the bwHPC initiative through the use of the JUSTUS HPC facility at Ulm University is acknowledged. This work contributes to the research performed at CELEST (Center for Electrochemical Energy Storage Ulm-Karlsruhe).

\bibliography{literatur_abbr}

\end{document}